\documentclass[twocolumn,aps,floats,floatfix,nofootinbib,prd,superscriptaddress,tightenlines]{revtex4}
\usepackage{graphicx}

\newcommand{\beq}{\begin{equation}}
\newcommand{\eeq}{\end{equation}}
\newcommand{\beqs}{\begin{eqnarray}}
\newcommand{\eeqs}{\end{eqnarray}}

\begin{document}
\title{Standard Model couplings and collider signatures of a light scalar}
\date{\today}

\author{JiJi Fan, Walter D. Goldberger, Andreas Ross,  and Witold Skiba }
\affiliation{Department of Physics, Sloane Laboratory, Yale
University, New Haven, CT 06520}

\begin{abstract}
The electroweak symmetry breaking (EWSB) sector of the Standard Model can
be far richer and more interesting than the usual single scalar doublet model. 
We explore scenarios where the EWSB sector is nearly scale
invariant and consequently gives rise to a light CP even scalar particle. 
The one-doublet SM is in that category, as are many other models
with either weakly or strongly coupled sectors that trigger EWSB\@. 
We study the couplings of the light scalar to the SM particles
that can arise from the explicit breaking of scale invariance
focusing on the possible differences with the minimal SM. 
The couplings of the light scalar to light fermions, as well 
as to the massless gauge bosons, can be significantly enhanced.
We find possible new discovery channels due to the decays
of the conformal scalar into $e^+e^-$ and $\mu^+\mu^-$ pairs as well as 
new production channels via light quark annihilation.

\end{abstract}
\maketitle

\section{introduction}

One of the main goals of experiments at the Tevatron and at the upcoming Large Hadron Collider (LHC) is to unravel the mechanism of electroweak symmetry breaking (EWSB).   The minimal way to obtain EWSB is to add to the Standard Model (SM) an $SU(2)$ doublet Higgs scalar field whose vacuum expectation value sets the electroweak scale $v\simeq 246~\mbox{GeV}$.    However, this simple model should perhaps be regarded as a parameterization rather than a dynamical explanation of EWSB.   In particular, the square of the Higgs mass parameter must be finely tuned (the hierarchy problem) and chosen to be of the correct sign in order to generate the scale $v$.

Resolving the shortcomings of the minimal Higgs picture has been the main motivation for constructing extensions of the SM in which new degrees of freedom arise at the TeV scale.  A common perception is that the close agreement between precision electroweak measurements~\cite{PDG} and the SM plus a light Higgs favors weakly coupled  TeV scale physics.  Extensions of this sort include supersymmetry~\cite{DG} or Little Higgs and related models~\cite{littleH1, littleH2,twinHiggs,Giudice:2007fh}.  However, the electroweak data does not necessarily exclude numerous models with strong dynamics, for example certain variants of technicolor~\cite{TC,walking}, or models with extra dimensions~\cite{RS,Csaki:2003zu,Agashe:2004rs,Gherghetta:2000kr}.   Such models might evade the experimental constraints if there are additional symmetries or if the predictions of the new strong interactions do not obey the naive dimensional analysis estimates.

Of course, it is also conceivable that the solution to the hierarchy problem will 
simply evade us at energies accessible to the LHC. Should this be the case, there is
still no guarantee that the EWSB sector is as simple as the minimal Higgs doublet, 
see for example Refs.~\cite{BahatTreidel:2006kx,Mantry:2007ar,Davoudiasl:2004aj}. 
More complicated scenarios that do not address the hierarchy problem
may seem unmotivated, but soon there will be data on the particle spectrum in the TeV range,
making theoretical prejudices unnecessary.

With so many dramatically different possibilities for the EWSB sector, 
it would be useful to have some sort of organizing principles for the new particle interactions. 
While we do not know of such principles in general, one may note that a large class of scenarios  contain in their spectrum a CP even scalar field that is light in comparison to the masses of all other new states.   This can be attributed to an enhanced symmetry, namely an (approximate) scale invariance which is spontaneously broken at a scale $f\geq v$.    The light scalar can be identified as the pseudo-Goldstone mode associated with the spontaneous breaking of conformal symmetry.    Its mass is naturally light,  proportional to the symmetry breaking scale $f$ times the small parameter that characterizes deviations from exact scale invariance.    We refer to such a pseudo-Goldstone boson as a \emph{dilaton}.

A particular example of this framework is the minimal $SU(2)$ doublet model itself.  In this case, the only sources of scale symmetry breaking are in the Higgs potential, and thus $f=v$ (we are neglecting the small  breaking of conformal symmetry due to SM loop effects in this discussion).   Choosing the mass parameter in the Higgs potential small compared to $v$, that is small explicit breaking of scale symmetry, results in a light Higgs which can be identified with the dilaton.    The underlying conformal invariance at the scale $v$ also governs all the properties of the  light Higgs particle, for instance tree-level couplings to SM fermions and gauge bosons.  Viewed in this light, Higgs-like phenomenology is a generic feature of theories that contain an approximate scale symmetry.   A similar pattern of couplings can be expected to arise in more complicated EWSB sectors with a light scalar field in their spectrum.

 Ref.~\cite{Goldberger:2007zk} considered the case where the dilaton comes from a strongly coupled nearly conformal extension of the Standard Model.   In such a scenario, most new states associated with EWSB are heavy and broad (masses and widths of order $\Lambda_{EW}\sim 4\pi v$), making them difficult to resolve at the LHC.     On the other hand, the theory has a light, narrow dilaton mode which could easily be mistaken for the ordinary SM Higgs boson. 
 
 In this paper  we consider the properties of a light scalar particle from a more general perspective.     In particular, we identify the new scalar resonance with the dilaton, and therefore use conformal symmetry to constrain its couplings to the SM.    Here, we do not necessarily assume that it is the only new particle observable at the LHC.  Of course, if other new states are discovered, that will certainly be crucial for distinguishing between different models.   In any case,  the symmetry arguments employed here would provide a useful way of parameterizing the properties of the lightest scalar particle in the spectrum.

For collider phenomenology, we concentrate on the operators that explicitly break scale invariance, as it is these terms that will be most useful for discrimination between the dilaton and a minimal Higgs scalar.   Of particular interest are operators that involve light SM fields in addition
to the dilaton, which are the most sensitive to the effects of symmetry breaking. 
In the SM, the dilaton-fermion couplings are proportional
to the fermion mass. This is true for the dilaton as well if there are no
explicit violations of scale symmetry. When scale symmetry is violated 
by operators involving fermions, the dilaton-fermion couplings need not
vanish in the massless fermion limit. An enhancement of such couplings is 
important for collider phenomenology, through new dilaton production
and decay channels that are negligibly small in the case of the SM Higgs boson. 

The paper is organized as follows. In Sec.~\ref{sec:setup}
we describe our setup.   There we use symmetry arguments to analyze the general properties of the dilaton couplings to SM fields.  In Sec.~\ref{sec:toy} we illustrate these ideas by showing how they are realized in perturbative models of extended Higgs sectors.   Our methods are most useful
when simple perturbative calculations are not available (that is when EWSB is induced by
spontaneous breaking of conformal symmetry due to strong interactions), but
the reasoning based on symmetry is clearly general.  Sec.~\ref{sec:colliders}
is devoted to phenomenology. We discuss constraints from the Tevatron
and predictions for the LHC\@. We conclude in Sec.~\ref{sec:conclusions}.

\section{Setup}
\label{sec:setup}

We are interested in describing the low-energy limit of SM extensions in which conformal invariance is spontaneously broken at some scale $f$ (or around $\Lambda_{CFT}\sim 4\pi f$ if the theory is strongly coupled).    We assume $f\geq v$,  and that any new particles with electroweak quantum numbers are heavy, with masses of order TeV.   Thus EWSB is parameterized in terms of the EW chiral Lagrangian, and we must include in our effective theory an $SU(2)$ matrix of EW Goldstone bosons $U(x)$.   The low-energy theory then consists of an electroweak singlet pseudo-Goldstone boson related to non-linearly realized scale symmetry (the dilaton mode) which is coupled to SM fields plus the chiral EW Lagrangian.

The couplings of the dilaton mode $\chi(x)$ to other light fields depend on how the SM is embedded in the conformal sector.   The most predictive situation (considered in Ref.~\cite{Goldberger:2007zk}, and reviewed below) corresponds to the case in which the SM gauge bosons and fermions arise as  composites of the spontaneously broken conformal field theory (CFT).   The couplings of $\chi(x)$ to the SM become model dependent if some of the SM fields are spectators to the strong CFT dynamics.    We consider this case in Sec.~\ref{sec:nonCFT}.

\subsection{Conformally embedded SM fields}

If the SM fields are embedded in the conformal sector, the prescription for coupling them to the dilaton is simple.   Start with the SM Lagrangian renormalized at some scale $\mu$,
\begin{equation}
\label{eq:SML}
{\cal L}_{SM} = \sum c_i(\mu) {\cal O}_i(x),
\end{equation}
where the operator ${\cal O}_i$ has definite scaling dimension $[{\cal O}_i]=d_i$, i.e. ${\cal O}_i(x) \rightarrow e^{\lambda d_i} {\cal O}_i( e^{\lambda} x)$ under the scale transformation $x^\mu\rightarrow e^{\lambda} x^\mu$.   Non-linearly realized scale invariance is then incorporated by introducing a flat direction (or ``conformal compensator field") $\chi(x)$ that transforms according to 
\begin{equation}
\chi(x) \rightarrow e^{\lambda} \chi(e^{\lambda} x)
\end{equation}
under scale transformations.    To make ${\cal L}_{SM}$ invariant under scale transformations one then replaces
\begin{equation}
\label{eq:cc}
c_i(\mu)\rightarrow \left({\chi\over f}\right)
^{4-d_i}  c_i\left(\mu {\chi\over f}\right)
\end{equation}
in Eq.~(\ref{eq:SML}).  Introducing a canonical kinetic operator for $\chi(x)$ and expanding about the VEV $\langle\chi\rangle =f$, one obtains to linear order in the fluctuation ${\bar\chi}=\chi-f$
\begin{equation}
\label{eq:lchi}
{\cal L}_\chi = - {\bar\chi\over f} {{T}^\mu}_\mu,
\end{equation}
 where $T^{\mu\nu}$ is the SM energy momentum tensor, and the trace includes quantum effects due to the conformal anomaly.   In addition, there are non-linear couplings of ${\bar\chi}$ to the SM, which are determined by scale invariance from Eq.~(\ref{eq:cc}), but these will not play a role in the collider phenomenology discussed later in this paper.

The above result gives, for example, tree level couplings of $\bar\chi$ to massive gauge bosons and fermions that are identical to those of the minimal Higgs, rescaled by a factor of $v/f$:
\begin{eqnarray}
\label{eq:SMchi}
\nonumber
{\cal L}_\chi &=& \left(\frac{2\bar{\chi}}{f}+\frac{\bar{\chi}^2}{f^2}\right)
   \left(m_W^2W_{\mu}^+W^{-\mu}+{1\over 2} m_Z^2 Z_\mu Z^\mu \right)\\
   & & {} - {\bar\chi\over f} \sum_\psi m_\psi {\bar\psi} \psi+\cdots.
\end{eqnarray}
This is consistent with the statement that, at tree level, the minimal Higgs boson is an example of a dilaton mode for which $v=f$.    

In addition there are loop induced couplings to massless gauge bosons, which are crucial for collider 
phenomenology.  At one-loop order in the gauge couplings and to all orders in $\chi$ these take the form.
\begin{equation}
\label{eq:gauge boson couplings}
{\cal L}_\chi = \left[\frac{\alpha_{EM}}{8 \pi
}b_{EM} (F_{\mu\nu})^2+\frac{\alpha_s}{8 \pi}b_{G}(G_{\mu\nu})^2\right]\ln{\chi\over f}.
\end{equation}
Here, $b_{EM}$, $b_G$ are one-loop beta function coefficients for the gauge couplings $e$ and $g_s$, including all CFT states heavier than the dilaton mass $m_\chi$ (once explicit symmetry breaking is included; see below).     By conformal invariance, the beta functions including all states (CFT plus SM) vanishes and we may trade $b_{EM}$, $b_G$ for minus the SM beta function coefficients with all relevant SM fields \emph{lighter} than $m_\chi$~\cite{Goldberger:2007zk}.   Thus, for example,
\begin{equation}
b_G= 11 - {2\over 3} n_f,
\end{equation}
where $n_f=6$ is the number of SM quarks.  In the case of the SM Higgs, loops of new heavy colored particles also induce operators such as those of Eq.~(\ref{eq:gauge boson couplings}) (e.g. from new chiral quarks) as well as dimension-six couplings proportional to $H^\dagger H (G_{\mu\nu})^2$.    Prospects for disentangling the contribution of both of these types of operators at the LHC are discussed in Ref.~\cite{Pierce:2006dh}.

So far our discussion has ignored terms that explicitly break conformal invariance.   Such terms must be present in order to generate a non-zero mass for the Goldstone boson $\chi$.   A simple way to do this is to deform the CFT by adding an operator of scaling dimension $\Delta\neq 4$,
\begin{equation}
\label{eq:deform}
{\cal L}_{CFT}\rightarrow {\cal L}_{CFT} + \lambda {\cal O}.
\end{equation}
This has the effect of generating non-derivative couplings for $\chi$ of the form
\begin{equation}
\label{eq:chipot}
V(\chi)=\chi^4 \sum^\infty_{n=1} c_n \left({\chi\over f}\right)^{n(\Delta-4)},
\end{equation}
which can be formally obtained by expanding the CFT formula for the vacuum energy $\langle 0| T \exp[i\int d^4 x \lambda {\cal O}(x)]|0\rangle$ to all orders in the symmetry breaking operator ${\cal O}(x)$.   The coefficients $c_n\sim\lambda^n$ in Eq.~(\ref{eq:chipot}) depend on unknown details of the dynamics in the underlying CFT.

It is not possible to use Eq.~(\ref{eq:chipot}) to make low energy predictions unless there exists a small expansion parameter.   We are interested in the case where the explicit breaking is small, which means that either the coupling $\lambda$ is small in units of $f$ (as would be the case for a light minimal SM Higgs), or because the operator ${\cal O}$ is nearly marginal, with scaling dimension $|\Delta-4|\ll 1$ (as in walking technicolor theories~~\cite{walking,Dietrich:2005jn}, or in Randall-Sundrum models with stabilization similar to that of~~\cite{Goldberger:1999uk}).    Both of these cases are discussed in more detail in Ref.~\cite{Goldberger:2007zk}.   We recall here that in both cases, to first order in the small breaking parameter,  the dilaton potential is completely determined in terms of the VEV $\langle\chi\rangle = f$ and the mass $m_\chi^2 = d^2 V/d\chi^2$.   Writing
\begin{equation}
V(\chi)= {1\over 2} m_\chi^2 \chi^2 + {g\over 3!}  {m^2_\chi\over f}   \chi^3 + \cdots,
\end{equation}
we find that $m_\chi^2$ is linearly proportional to the symmetry breaking parameter, $m_\chi^2 \sim \epsilon f^2$, and therefore $\chi$ is naturally light relative to the scale $f$.    From now on, we will take the dilaton mass as the measure for the size of symmetry breaking and thus we define $\epsilon\equiv m^2_\chi/f^2\ll1$.   In addition, the cubic coupling is given to leading order in the symmetry breaking by
\begin{equation}
\label{eq:cubic}
g=\left\{\begin{array}{ll}
\Delta+1 & \rm{for~{\hat \lambda}\ll 1},\\
5 & \rm{for}~|\Delta-4|\ll 1.
\end{array}\right.
\end{equation}
(Here ${\hat\lambda}\propto\lambda$ is a dimensionless measure of the coupling $\lambda$).  Note in particular, that for the Higgs case, with $\Delta=2$, this agrees with the standard result on the cubic coupling.  The nearly marginal case can be realized, for example, by the Coleman-Weinberg effective potential for the Higgs~\cite{Gildener:1976ih}, see also Ref.~\cite{Noble:2007kk}. All other non-derivative self-couplings are also determined in terms of $m_\chi$ and $f$, at least to leading order in the symmetry breaking parameter.

The effects of symmetry breaking also modify the couplings of $\chi$ to the SM fields.  In the symmetry limit the couplings are given in Eq.~(\ref{eq:SMchi}).   Because of symmetry breaking, after EWSB the coupling becomes
\begin{equation}
{\cal L}_Y  =- {\chi\over f} \sum_\psi  m(z){\bar\psi}{\psi},
\end{equation}
where $m(z)$ is an analytic function of $z= {\hat\lambda}(\chi/f)^{\Delta-4}$.   In general, $m(z)$ is a matrix in fermion flavor space (see below) and reproduces the effects of an arbitrary number of insertions of the symmetry breaking perturbation $\lambda\int d^4 x {\cal O}(x)$ into amplitudes.  This gives to linear order in ${\bar\chi}$,
\begin{equation}
{\cal L}_Y = -\sum_\psi m_\psi {\bar\psi} {\psi} - {\bar\chi} \sum_\psi Y_\psi {\bar\psi} {\psi}+\cdots,
\end{equation}
where in the ${\hat \lambda}\ll 1$ case, one gets
\begin{eqnarray}
\label{eq:mandy}
\nonumber
m_\psi &=& m(0)+ {\hat \lambda} m'(0),\\
Y_\psi &=& {1\over f}\left[m(0) +{\hat \lambda} (\Delta-3) m'(0)\right],
\end{eqnarray}
while for $|{\Delta}-4|\ll1$, we find $m_\psi= m({\hat \lambda})$ and 
\begin{equation}
\label{eq:mandyeps}
Y_\psi  = {1\over f}\left[m_\psi+ (\Delta-4) m'({\hat \lambda})\right].
\end{equation}
Due to symmetry breaking, the Yukawa couplings to $\chi$ consist both of a term proportional to the fermion mass $m_\psi$ and a mass independent term.   While this additional mass independent correction is unimportant for heavy fermions, it may in fact dominate for fermions with $m_\psi\ll f$.   This can lead to interesting physical consequences, for example novel production channels at hadron colliders, $u\bar{u}\to\chi, d\bar{d}\to\chi$, and unexpected discovery channels, $\chi\to e^+e^-, \chi \to \mu^+\mu^-$, which we will explore in detail below.   Note that this effect is expected to be more natural in the case of symmetry breaking by a marginal operator, as achieving $Y_\psi\gg m_\psi/f$ in the case ${\hat \lambda}\ll 1$ requires a fine tuning in Eqs.~(\ref{eq:mandy}) between terms that are formally of different orders in the expansion.   We discuss an explicit model of this type in the next section.

In this discussion, we have neglected the flavor structure of the
symmetry breaking couplings. In fact, each term in the expansion of the function $m$ is an independent $3\times3$ matrix in flavor space.   Thus the
usual bi-unitary transformations that diagonalize the fermion
mass matrices will not necessarily diagonalize the scale symmetry breaking part of the fermion Yukawas unless one assumes a
particular flavor ansatz, e.g., minimal flavor
violation~\cite{Ali:1999we}. The off-diagonal
couplings would mediate flavor changing processes like $\mu\to3e$
and are constrained by experiments.  For the purposes of this
paper, we will neglect this issue and assume that the
couplings are diagonal in flavor space but do not necessarily have
the same hierarchy structure as the SM Yukawa matrices.

The presence of operators that violate the chiral symmetry,
like the dilaton coupling to the fermions, spoils the
technical naturalness of fermion masses. 
Radiative corrections can generate sizable
contributions to the mass term.   Defining $\epsilon\delta Y= (\Delta-4) m'({\hat \lambda})/f$ in Eq.~(\ref{eq:mandyeps}), there is an interaction of two dilatons with fermions, $\epsilon\delta Y \bar{\chi}^2 \bar{\psi}\psi/f$ that generates a radiative  correction to $m_{\psi}$ that is of order
\begin{equation}
\delta m_\psi\sim \frac{\Lambda^2}{(4 \pi f)^2}  \epsilon\delta Y f.
\end{equation}
By setting the ultraviolet cutoff $\Lambda\sim 4\pi f$ this becomes $\delta m_\psi\sim \epsilon\delta Y f$. 
The amount of fine tuning depends on $\epsilon$, with less fine tuning required when $\epsilon$
is small.

For instance, for the muon, lack of fine tuning would imply $  \epsilon\delta Y_\mu \approx 10^{-4}$.
We will consider larger values of $\epsilon \delta Y$ at the expense of possible fine tuning in the underlying theory.  It is useful to compare this situation to that in the SM\@, where the naturalness of 
small fermion masses can also be upset by new physics. For example, heavy states beyond the SM can generate chirality-violating higher dimensional operators such as
\begin{equation}
{c_1\over \Lambda^2}\bar{L} H \sigma^{\mu\nu}eB_{\mu\nu},  \,\,{c_2\over\Lambda^2} \bar{L} H
\sigma^{\mu\nu}eW_{\mu\nu}.
\end{equation}
These contribute an amount $\delta m_e\sim c_i e v/(16\pi^2)$  to the electron mass through radiative corrections. One linear combination of these operators also contributes to the magnetic moment and is therefore small, but the orthogonal combination is not 
tightly constrained.  In any case, there are other operators,
for example $\bar{L} (D_\mu H) D^\mu e$, that are unconstrained and can also radiatively generate contributions to the electron mass.  
In theories with strong interactions or extra dimensions, one expects that the coefficients $c_i$ can be of order one, or perhaps larger if there are $N\geq 1$ species of heavy states that contribute.
Consequently, the contribution to the electron mass could be as large as 
$\delta m_e \sim N v/(16\pi^2)$. 

\subsection{SM fields as spectators}
\label{sec:nonCFT}

It is also possible for the SM fields to be realized as weakly coupled spectators to the CFT dynamics.  In this case, the theory above the scale $f$ consists of the spectator SM fields coupled to CFT operators.   The role of EWSB in scenarios with SM spectators coupled to a CFT sector has been studied in Refs.~\cite{Delgado:2007dx,Delgado:2008rq,Lee:2008ph}.

Generically one expects all couplings between the SM and the CFT that are allowed by the gauge symmetries to be present.     For illustration we will consider the example
\begin{equation}
\label{eq:SMspec}
{\cal L}_F = {\cal O}(x) {\bar \psi} \psi(x),
\end{equation} 
with $\psi$ an SM fermion and ${\cal O}(x)$ a CFT operator of engineering (mass) dimension one and arbitrary scaling dimension $\Delta$.   For simplicity, we have suppressed EW quantum numbers.  

In the effective theory below the scale $f$, Eq.~(\ref{eq:SMspec}) induces both a mass term for $\psi$ and interactions of $\psi$ with any number of scale Goldstone modes.   For vanishing dilaton momenta, all such couplings are related to the fermion mass, $m_\psi=\langle 0|{\cal O}| 0 \rangle$.  E.g., the one-dilaton coupling is fixed by the matrix element
\begin{equation}
\langle \chi(p^\mu\rightarrow 0)| {\cal O}(0)|0\rangle = -{i\over f}  \langle 0| [ Q_S, {\cal O}(0)]|0\rangle = {m_\psi\over f} \Delta,
\end{equation}
where $Q_S=\int d^3 {\bf x}\, S^0({\bf x},0)$ is the charge generated by the scale current $S^\mu = x^\nu {T^\mu}_\nu$ of the CFT.   More generally, the low-energy theory contains a coupling
\begin{equation}
\label{eq:yuk}
{\cal L}_\psi =  - m_\psi \left({\chi\over f}\right)^\Delta {\bar\psi} \psi
\end{equation}
that includes the Yukawa $Y= (m_\psi/f)\Delta$ after expanding about the VEV.   

Thus, under the assumption that couplings induced by more insertions of the operator Eq.~(\ref{eq:SMspec}) are subleading, one finds that all dilaton couplings to $\psi$ are proportional to the fermion mass $m_\psi$.   This result holds quite generally:      any SM field that acquires a mass term after EWSB has dilaton couplings proportional to its mass.    The proportionality constant is the dimension of the CFT operator that couples to the SM field.   As shown in~\cite{Csaki:2007ns} this result also holds if the symmetries allow for couplings that induce mixing between the SM spectators and the CFT.

It is important to note that the result in Eq.~(\ref{eq:yuk}) holds only if multiple insertions of Eq.~(\ref{eq:SMspec}) are somehow suppressed.     If this is not the case, then there will be multiple independent contributions to the parameters $m_\psi$ and the dilaton Yukawa, spoiling any correlation between them.    Also, while couplings such as that in Eq.~(\ref{eq:SMspec}) break conformal symmetry explicitly, it may still be necessary to deform the CFT as in Eq.~(\ref{eq:deform}) in order to stabilize the dilaton.   Again, this will have the effect of inducing dilaton couplings which are not necessarily proportional to mass.

If the strong and electroweak gauge interactions do not emerge from the conformal sector, the loop induced couplings $\chi gg,$ $\chi \gamma\gamma$ become model dependent.   This is true even for the SM Higgs in the presence of heavy states associated with new physics.    We parametrize our ignorance regarding this UV physics in terms of the couplings in Eq.~(\ref{eq:gauge boson couplings}), with $b_{EM}$, $b_G$ replaced by arbitrary coefficients $c_{EM}$, $c_G$.   Note that we take $c_{EM}$, $c_G$ to include only the contributions from heavy (non-SM) states.    The contribution of SM particles to $\chi\rightarrow gg$, $\chi\rightarrow \gamma\gamma$ must be explicitly added by computing triangle diagrams.

\section{Toy Examples}
\label{sec:toy}
We would like to illustrate some of the observations from the previous section in the context of
two perturbative examples. First, we will consider a model with a Higgs doublet
and a singlet. We will choose the parameters such that $f \gg v$ and the lightest 
scalar, which is a linear combination of the Higgs field and the singlet,  is mostly the singlet state.
We can easily calculate the cubic coupling of such scalar using the analysis in Sec.~\ref{sec:setup}.
Our second example involves two Higgs doublets. We show that the Yukawa couplings
of the light scalar do not necessarily vanish in the limit of zero fermion masses. 

In the first example, we will concentrate on the scalar potential only, as other interactions are not relevant
for our analysis. We consider an electroweak doublet $H$ and a real singlet $S$. The scalar 
potential is 
\beq
 V(H,S)=\frac{\lambda}{2} \left(H^\dagger H -\frac{\alpha^2}{2} S^2\right)^2 + \frac{\zeta}{4} \left(S^2 -\mu^2\right)^2.
 \label{eq:example1}
\eeq
A more realistic version of this toy model and its connection to the simultaneous breaking of EW and conformal symmetries has been recently studied in~\cite{Foot:2007as}.

The mass term for $S$ breaks the conformal symmetry explicitly, and to make that breaking small
we assume that $\zeta \ll 1$. This ensures that the mass term is small compared
to the expectation value of $S$.     Note that in the formal limit $\zeta\rightarrow 0$, the field $S$ becomes a flat direction, which can be identified with the dilaton of the previous section.  When  $\lambda\rightarrow\infty$, the fluctuations of $H$ about its VEV, $\langle H^\dagger H\rangle = \alpha^2 \langle S^2\rangle/2$, are infinitely massive and can be integrated out.   The resulting low-energy effective theory is precisely the EW chiral Lagrangian for the Goldstone components of the Higgs field $H$ coupled to an electroweak singlet dilaton mode proportional to the flat direction $S$.

Approximate scale invariance can still be used to understand the low energy dynamics away from the limit $\zeta\rightarrow 0$, $\lambda\rightarrow\infty$.    At the minimum of the potential,  $\langle S \rangle =\mu$ and  $\langle H^T \rangle = (0, v/\sqrt{2})$, with $v=\alpha \mu$. After EWSB there are two physical scalars, which we denote as $\chi$ and $\phi$. It is straightforward to express the fluctuations around the corresponding minima $S= \mu +s$ and $ H^T = (0, {v+ h\over \sqrt{2}})$
 in terms of the mass eigenstates $\chi$ and $\phi$
 \begin{equation}
\label{eq:chiandphi}
\left( \begin{array}{c}
h\\
s
\end{array}\right) = 
\left(\begin{array}{cc}
\cos\theta  & \sin\theta\\
-\sin\theta  & \cos\theta
\end{array}\right) \cdot
\left( \begin{array}{c}
\phi\\
\chi
\end{array}\right),
\end{equation}
where in terms of the parameters in $V(H,S)$ 
\begin{equation}
\tan 2 \theta ={\alpha \lambda v^2\over \lambda v^2 (1-\alpha^2)/2 -\zeta \mu^2}.
\end{equation}
We choose the parameters such that $\chi$ is the lighter  mass eigenstate.
 It is clear that (in the notation of the previous section) $v/f=\sin\theta$, since all tree-level couplings of $\chi$
 to the SM fields are those of the Higgs boson multiplied by $v/f$. For $f \gg v$ the light
 eigenstate is predominantly the singlet field $s$, yet its couplings are Higgs-like due to
 the underlying conformal symmetry.
 
  We can now write the potential in Eq.~(\ref{eq:example1})
 in terms of the mass eigenstates in Eq.~(\ref{eq:chiandphi}) 
 and then calculate the masses and couplings of the physical states. The expression is too long
 to be reproduced here.   It is interesting to note, however, that the cubic coupling of $\chi$ has a very simple form when expressed in terms of its mass and the scale $f$ which governs its couplings to SM fields, 
  \beq
   V={1\over 2} m_\chi^2 \chi^2 +{1\over 3!}  \frac{3 m_\chi^2}{f} \chi^3 + \ldots,
   \label{eq:ex1potential}
\eeq 
where we carried out this calculation to first order in $\zeta\ll 1$. 

The same result could have been obtained without explicit reference to the scalar potential.    All that matters is that for energies below the mass of the heavy scalar $\phi$, the theory exhibits an approximate scale invariance.  This scale symmetry plus the (small) sources of symmetry breaking completely fix the properties of the light mode $\chi$.   Indeed, given that the largest source of symmetry breaking in Eq.~(\ref{eq:example1}) is an operator of scaling dimension $\Delta=2$, the result of Eq.~(\ref{eq:ex1potential}) can be obtained quite easily, without explicit calculation. In particular, Eq.~(\ref{eq:cubic})  shows that the dilaton cubic coupling
is $(\Delta+1) m_\chi^2/f$, in agreement with Eq.~(\ref{eq:ex1potential}).   What is interesting here is that this result is universal.   None of the details of the underlying model enter into its derivation, save for the fact that the lightest scalar is a dilaton.

We now turn to a discussion of how scale symmetry breaking affects the couplings to the SM fields.   In particular we will consider fermion Yukawa couplings.   Our example is essentially a model discussed in Refs.~\cite{Mantry:2007ar,Mantry:2007sj,Randall:2007as}, but it would be easy to generate similar behavior with more scalar fields.
 Consider a two Higgs doublet model where both doublets couple to the same type
of fermions, for example the down quarks
\beq
 {\mathcal L}_Y=-y_1 \bar{Q} H_1 d - y_2 \bar{Q} H_2 d + \mbox{h.c.}.
 \label{eq:ex2yukawa}
\eeq
We are not concerned with possible flavor-changing neutral currents in such
a scenario~\cite{GlashowWeinberg} as our example is an illustration of principle only and
 its parameters will have to be fine-tuned anyway.  The potential for the two doublets is taken to be
 \beqs
  V(H_1,H_2) & = & \frac{\lambda}{2} \left(H_1^\dagger H_1 -\frac{\mu^2}{2}\right)^2 + M^2 H_2^\dagger H_2  \nonumber \\
    & &{} -  \left(g  H_2^\dagger H_1 H_1^\dagger H_1 + \mbox{h.c.}\right) + \ldots, 
    \label{eq:example2}
\eeqs
where we have omitted other quartic terms that are not relevant for our argument.
For large values of $M$ we can integrate out the doublet $H_2$ by substituting
$H_2 \rightarrow g H_1 H_1^\dagger H_1/M^2$. 

We expand around the minimum of
the potential where the lower component of $H_1$ is approximately $(v + h)/\sqrt{2}$.
The Yukawa terms in Eq.~(\ref{eq:ex2yukawa}) become
\beq
 {\mathcal L}_Y=  -y_1 \bar{d}\,\left[ \frac{v + h}{\sqrt{2}}\right] d -
     y_2\bar{d}\,  \frac{g}{M^2}\left[\frac{v+ h}{\sqrt{2}}\right]^3 d + \mbox{h.c}.
\eeq
Therefore the mass of the $d$ quark is
\beq
  m_d= \frac{v}{\sqrt{2}} \left( y_1 +  \frac{ g v^2}{2 M^2} y_2 \right),
\eeq
while its Yukawa coupling is
\beq
  y_d= \frac{1}{\sqrt{2}} \left( y_1 + {3 g v^2\over 2 M^2} y_2\right).  
\eeq 
These results agree with the general argument leading to Eq.~(\ref{eq:mandy}), where we identify the relevant source of scale symmetry breaking with the dimension-six operator $ (H^\dagger_1 H_1)^3$ induced by integrating out $H_2$ through its equation of motion.

The original Yukawa couplings, $y_1$ and $y_2$, enter the equations above in two
independent linear combinations. For light fermions, it is possible to balance the two terms
such that $m_d \rightarrow 0$ while $y_d\rightarrow -y_1\sqrt{2}$.   In this limit,  the Yukawa coupling of the light fermion can be significantly enhanced compared to the case of the SM with a Higgs doublet,  where the Yukawa coupling is proportional to fermion mass.   Of course, this limit has been achieved by fine tuning, which may be less severe in models with large couplings.    We may also find out experimentally that the there are fine tunings in the fermion sector and the Yukawas of the light fermions are much larger than naively expected.

\section{Collider Phenomenology}
\label{sec:colliders}

We parameterize the light dilaton couplings that are relevant for collider physics in terms of the following effective Lagrangian, 
\begin{eqnarray}
\label{eq:effL}
\nonumber
{\cal{L}}_{\chi}&=&-\frac{\bar{\chi}}{f}\sum_{\psi}\left(m_\psi+\epsilon y_{\psi}^{(1)}v\right)\,\bar{\psi}\psi \nonumber \\
&&{} +\left(\frac{2\bar{\chi}}{f}+\frac{\bar{\chi}^2}{f^2}\right)
   \left(m_W^2W_{\mu}^+W^{-\mu}+{1\over 2}m_Z^2 Z_\mu
Z^\mu\right) \nonumber \\
&&{} +{\bar\chi\over f}\left[\frac{\alpha_{EM}}{8 \pi
}c_{EM} (F_{\mu\nu})^2+\frac{\alpha_s}{8 \pi
}c_{G} (G_{\mu\nu})^2\right],
\end{eqnarray}
where the fluctuations about the VEV $\langle\chi\rangle=f$ are given in terms of a canonically normalized field $\bar\chi$ with mass $m_\chi$.    The constant $\epsilon \equiv m^2_\chi/f^2$ parameterizes the size of deviations from exact scale invariance.   As discussed in Sec.~\ref{sec:setup}, the couplings to massless gauge bosons, encoded in the coefficients $c_{EM}$ and $c_G,$ depend sensitively on assumptions about ultraviolet physics.   In particular, if it is assumed that the electroweak and QCD couplings become asymptotically conformal, then  $c_{EM}, c_G$ are the respective one-loop beta function coefficients, calculated including all relevant fields lighter than $m_\chi$.

We will investigate the properties of the dilaton mode at
colliders as a function of  the model parameters $m_\chi,f$, the
nine additional contributions to the Yukawa couplings $y_\psi^{(1)},$ and the dilaton couplings to the massless gauge bosons $c_{EM}$ and $c_G$. 
Instead of sampling the vast multi-dimensional parameter space, we will only
study several representative regions where the dilaton properties can be significantly 
modified relative to those of the SM Higgs boson. See Ref.~\cite{Duhrssen:2004cv} 
for an overview of the prospects for the determination of Higgs couplings at the LHC\@.

\subsection{New Production Channels}

For the minimal SM Higgs boson, the main production mechanisms at a hadron collider are the gluon-gluon fusion mechanism
$gg\to\chi$, associated production with $W/Z$ bosons
$q\bar{q}\to V+\chi$, weak vector boson fusion processes
$qq\to V^* V^* \to qq+\chi$, and associated production with
a top quark pair,  $gg,q\bar{q}\to
t\bar{t}+\chi$~\cite{Djouadi:2005gi}.    These are also the dominant channels for the case of a light dilaton.

For a light dilaton, the cross sections of the
tree-level processes scale as $(v/f)^2$ compared to the SM Higgs
production cross sections. The gluon fusion mechanism, on the other hand,
depends crucially on the coupling $c_G$. For instance, embedding
QCD in the conformal sector will result in an enhancement in
$c_G$ which may overcome the scaling suppression factor
$(v/f)^2$ in the cross section.

\begin{figure}[h]
\begin{center}
\includegraphics[width=1\linewidth]{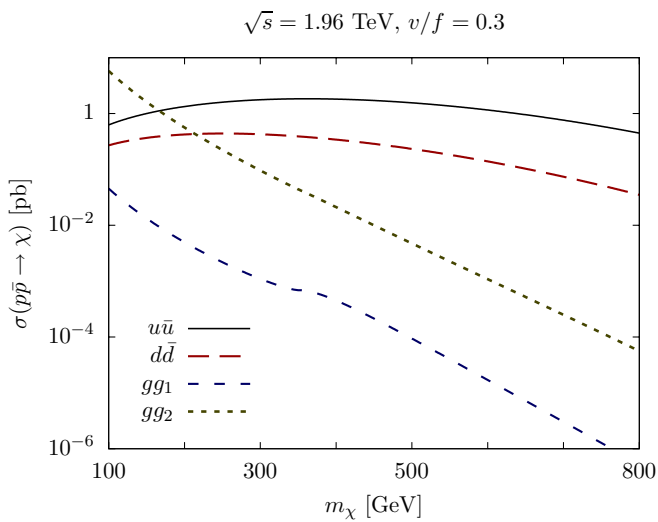}
\caption{Tevatron production cross sections for different channels at LO as a function of $m_\chi$. Shown are the cross sections $u\bar{u} \to \chi$ with $y^{(1)}_{u}=1$,  $d\bar{d} \to \chi$ with $y^{(1)}_{d}=1$, and $gg \to \chi$  for two different coupling values. The  $gg_1$ curve corresponds to a SM like $gg\chi$ coupling arising solely from the top quark loop while the $gg_2$ curve 
corresponds to the scenario where the SM is fully embedded in the CFT such that $c_G=11-2 n_f/3 $. \label{fig:production}}
\end{center}
\end{figure}

\begin{figure}[h]
\begin{center}
\includegraphics[width=1\linewidth]{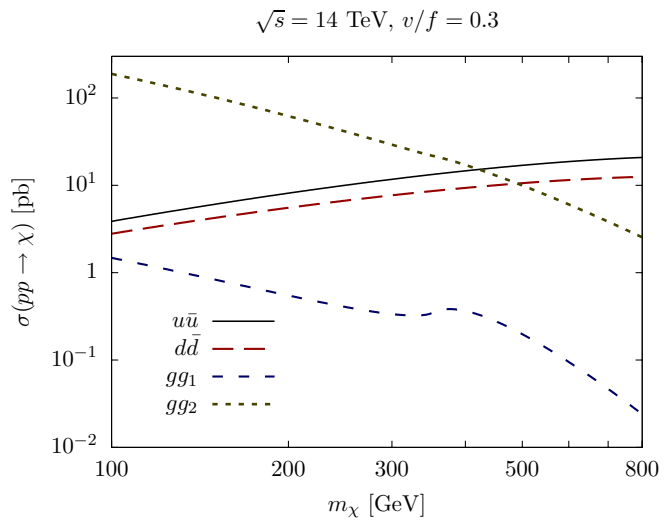}
\caption{LHC production cross sections for different channels
 at LO as a function of $m_\chi$. The correspondence between line types and processes is
 as in Fig.~\ref{fig:production}.
 \label{fig:production2}}
\end{center}
\end{figure}

A possible enhancement of the couplings between $\chi$ and the first
generation quarks opens the new production channels
$u\bar{u}\to\chi$, $d\bar{d}\to\chi$ at the LHC and the 
Tevatron. For example, at the LHC the tree-level cross section for these processes (employing the narrow width approximation) is 
\begin{equation}
\sigma_{LO}(pp\to \chi)=\sigma_0^\chi
\tau_{\chi}\frac{d{\cal{L}}^{q\bar{q}}}{d\tau_{\chi}},
\end{equation}
where 
\begin{equation}
\sigma_0^\chi=\frac{\pi v^2 m_\chi^2}{3f^6}   (y_q^{(1)})^2,
\end{equation}
and
\begin{equation}
\frac{d{\cal{L}}^{q\bar{q}}}{d\tau_{\chi}}=\int_{\tau_{\chi}}^1{\frac{dx}{x}f_q(x,\mu_F^2)f_{\bar{q}}(\frac{m_\chi^2}{xs},\mu_F^2)}.
\end{equation}
Here, $f_q,f_{\bar{q}}$ are the parton distribution
functions for $q=u,d$, and $\tau_\chi=m_\chi^2/s$, with $s$ the collider center of mass
energy squared, is the usual Drell-Yan variable.
In our plots below, we take $y^{(1)}_u,y^{(1)}_d=1$ and $v/f=0.3$. We use
the CTEQ5 parameterization for the parton distribution functions and choose the
factorization scales to be $\mu_F=m_\chi$. The total hadronic cross
sections at leading order (LO) are are shown in
Fig.~\ref{fig:production} and Fig.~\ref{fig:production2} for
the Tevatron and the LHC, respectively.

Given the choices of parameters in Fig.~\ref{fig:production}, the cross sections of the light quark annihilation channels are roughly $1\, \rm{pb}$ at the Tevatron over the entire range of masses.  The quark annihilation cross sections exceed the gluon fusion channel for $m_\chi>150\,
\rm{GeV}$, even when the  $gg\chi$ coupling constant is enhanced to $c_G=11-2 n_f/3$. At LHC energies, the $q{\bar q}\rightarrow \chi$  cross sections increase with increasing $m_\chi$ for reasonable values of the mass. The sum of the $u,d$ channels is about $10~\rm{pb}$ for
$m_\chi \sim 120~\rm{GeV}$ and about $20~\rm{pb}$ for $m_\chi
\sim 700~\rm{GeV}$. 

The growth of the cross section with mass is due to the symmetry breaking Yukawa couplings in Eq.~(\ref{eq:effL}), which increase with $m^2_\chi$ for fixed $f$.    This is, of course, very different than for processes mediated by dilaton couplings that are already present in the symmetry limit $\epsilon\rightarrow 0$.     Such couplings are independent of $m_\chi$ and the relevant cross sections decrease rapidly as $m_\chi$ increases.    For large $m_\chi$, the $u\bar{u},d\bar{d}$
channels become comparable to the $gg$ channel with $c_G=11-2 n_f/3$ and surpass it for $m_\chi> 350~\rm{GeV}$.

Another possible production mechanism resulting from  enhanced
$u\bar{u}\chi$, $d\bar{d}\chi$ couplings is dilaton
bremsstrahlung associated with one or two jets, e.g.,
$ug\to u^* \to u\chi$. We have checked that at the LHC, the cross
section of bremsstrahlung associated with one jet is
 $\sim 1~\rm{pb}$ but still smaller than that of the $q{\bar q}\rightarrow\chi$ channel.

\subsection{New Discovery Channels}

For a dilaton with Higgs-like couplings, the dominant decay channel is
$\chi \to b \bar{b}$ in the low mass region and $\chi \to WW/ZZ$
in the high mass region. Just as for the SM Higgs, different strategies must be employed in order to extract the dilaton signal  depending on the precise value of $m_\chi$ and the dominant production mechanism. 
For example, the most promising discovery mode for the
gluon fusion channel is $\chi \to ZZ \to 4l$ or $\chi \to WW
\to ll\nu \nu$ if $\chi$ is sufficiently heavy. As we discussed earlier, new
heavy states can modify the couplings  $gg \chi$ and $\gamma
\gamma \chi$, as is also the case for the minimal SM Higgs couplings if there is new physics~\cite{MW,Hankele:2006ma}.   The $\chi\to \gamma \gamma$
process can be very important in the SM. It is likely that the $\gamma
\gamma \chi$ coupling will be suppressed compared to the SM
$\gamma \gamma h$ coupling~\cite{Goldberger:2007zk}, thus we do not discuss this channel further.

The possibility of enhanced Yukawa couplings to light SM fermions would
imply new discovery channels that are not significant in the case of the SM Higgs.
For example, an enhancement in the $\chi \ell{\bar\ell}$ vertex, with $\ell=e,\mu$ would result in experimentally clean new discovery signatures $\chi \to e^+e^-,\chi \to
\mu^+\mu^-$ which are negligible for the SM Higgs 
due to the small Yukawa couplings.   

For illustration, we consider a point in parameter space with an ``anomalous" Yukawa coupling to muons, but with all other light fermion Yukawas proportional to their mass (in the notation of Eq.~(\ref{eq:effL}), we set  $y^{(1)}_{\psi}=0$ except for $\psi=\mu$).   In this case we can compare with
the existing studies of $h \to \mu^+\mu^-$ via gluon
fusion~\cite{Han:2002gp,Cranmer:2006zs}, see also \cite{Plehn:2001qg}. 
In these references, it is shown that for the minimal Higgs,
$gg\to h \to \mu^+\mu^-$ is rather clean but requires a
large integrated luminosity of ${\cal{L}}=300~\rm{fb}^{-1}$ for
a $3 \sigma$ signal in the mass range $m_h \sim 115-140~\rm{GeV}$.

In the dilaton case, the enhancement of this process can be dramatic. The cross
section for the $gg \to \chi \to \mu^+\mu^-$ is
proportional to $\sigma(gg \to \chi)\times\mbox{BR}(\chi \to
\mu^+\mu^-)$. The cross section $\sigma(gg \to \chi)$, as we
discussed above, is sensitive to contributions from physics beyond the SM.   Due to the enhanced $\chi\mu^+\mu^{-}$ coupling, the branching fraction $\mbox{BR}(\chi \to \mu^+\mu^-)$ may also deviate significantly from that of the minimal Higgs, $\mbox{BR}(h \to \mu^+\mu^-)\sim 10^{-4}$ in the  $m_h \sim 115-140~\rm{GeV}$ mass range. 

\begin{figure}[h]
\begin{center}
\includegraphics[width=1\linewidth]{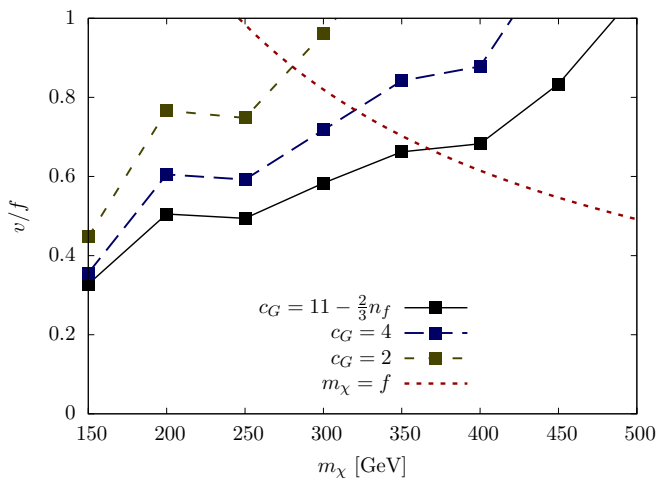}
\caption{Upper limits on $v/f$ as a function of $m_\chi$ implied by the Tevatron lepton pair data.   We assume $y^{(1)}_\mu=1$ and all other $y^{(1)}_{\psi}=0$.   The different curves correspond to bounds for several choices of the $\chi gg$ coupling $c_G$. Also shown is the
line $m^2=f^2$ beyond which approximate scale symmetry no longer holds. \label{fig:tevee}}
\end{center}
\end{figure}

\begin{figure}[h]
\begin{center}
\includegraphics[width=\linewidth]{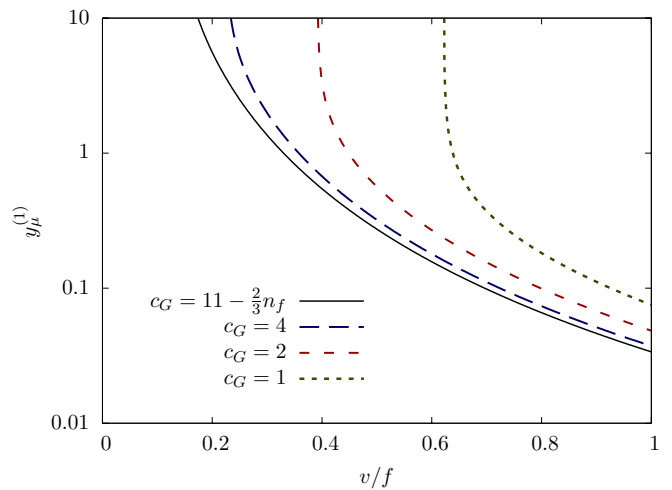}
\caption{Upper limits on $y_\mu^{(1)}$ as a function of $v/f$ implied by the Tevatron lepton pair data.   We assume $m_\chi=150~\rm{GeV}$, $y^{(1)}_\mu\neq 0$, and all other
$y^{(1)}_{\psi}=0$.   From left to right, $c_G=11-2/3\, n_{f},4,2,1$.
\label{fig:tevee2}}
\end{center}
\end{figure}

The  $gg \to \chi \to \mu^+\mu^-$ channel is potentially accessible at the
Tevatron.  Therefore, the null search result for new scalar particles
decaying into final states containing muon pairs ~\cite{Abulencia:2005nf} places limits on the model
parameters $m_\chi$, $f$, and $y_\mu^{(1)}$.  Using
the $95 \%$ upper limits on $\sigma(p \bar{p} \to \chi)\times \mbox{BR}(\chi
\to \mu^+\mu^-)$ for several values of  $m_\chi$ in the range $150~\rm{GeV}$ to
$800~\rm{GeV}$ reported in Ref.~\cite{Abulencia:2005nf}, we present the constraints
on the parameters $v/f$ and $y_\mu^{(1)}$ in Fig.~\ref{fig:tevee}
and Fig.~\ref{fig:tevee2}.  In these plots we assume that only the coupling
to muons is enhanced, with the couplings to other light fermions suppressed by their mass.

Fig.~\ref{fig:tevee} shows the limit on $v/f$ as a function of
$m_\chi$, where we set $y_\mu^{(1)}=1$.   The most stringent limits on $v/f$ correspond to large $\sigma(gg\rightarrow \chi)$ (large $c_G$), while the bound on $v/f$ relaxes as $m_\chi$ increases. There are two effects responsible for this:   the decrease in  $\sigma(gg\rightarrow\chi)$ for large masses, and $\mbox{BR}(\chi\rightarrow \mu^{+}\mu^{-})$ becomes negligible compared to the $WW, ZZ$ channels as
$m_\chi$ increases. 
We do not consider $m_\chi>f$  as this corresponds to badly broken scale symmetry.
The boundary $m_{\chi}\leq f$ is also indicated in Fig.~\ref{fig:tevee}.
Fig.~\ref{fig:tevee2} illustrates the bounds for $m_\chi=150~{\rm GeV}$ expressed as 
a limit on $y^{(1)}_\mu$ as a function of $v/f$.

\begin{figure}[h]
\begin{center}
\includegraphics[width=1\linewidth]{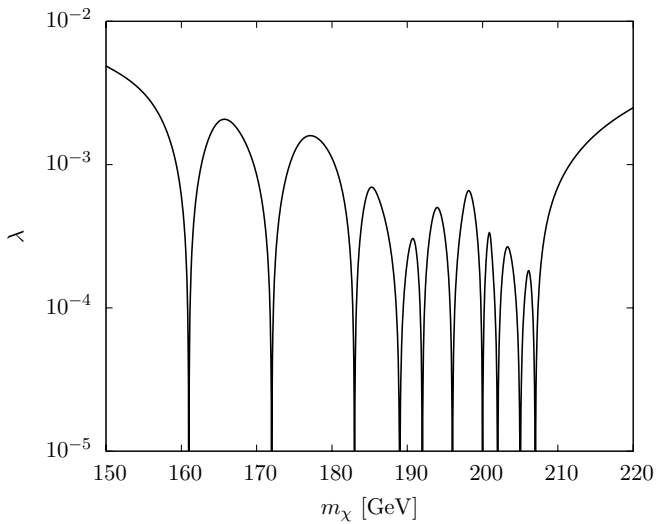}
\caption{$95\%$ CL upper limit on the product $\lambda=(v/f)^2  y_e^{(1)}  y_\mu^{(1)}$ implied by the LEP2 data, as a function of $m_\chi$.   The plot only shows a small range
of $m_\chi$ around the energies for which LEP2 collected data, 
where the bounds are most stringent. \label{fig:lep2}}
\end{center}
\end{figure}

A scalar with enhanced couplings to leptons could have been observed at LEP\@. The absence of significant deviations from the SM predictions at LEP implies constraints on such couplings.
As an example, we consider the LEP2 data in the $e^+e^- \to  \mu^+ \mu^-$ channel~\cite{LEP}, which constrains the product $y_e^{(1)}  y_\mu^{(1)}$.    The bounds are presented
in Fig.~\ref{fig:lep2}.   These bounds are very stringent at the center of mass energies
where LEP collected data. They are correspondingly weak away from the LEP2 mass range.

\begin{figure}[h]
\begin{center}
\includegraphics[width=1\linewidth]{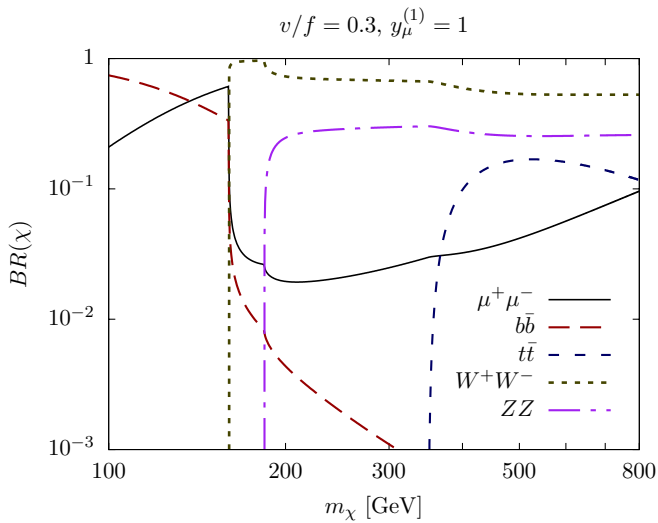}
\caption{Dilaton branching ratios as a function of $m_\chi$ for $v/f=0.3,$ $c_G=1$\label{fig:br}.   We have set $y^{(1)}_\mu=1$ and $y^{(1)}_\psi=0$ for all other fermions.}
\end{center}
\end{figure}

In Fig.~\ref{fig:br} we display the dilaton branching fractions in the case where only the muon couplings are enhanced.  We choose $v/f=0.3$ and $y^{(1)}_\mu=1$.   As for the minimal SM Higgs boson, for  $m_\chi>160~\rm{GeV}$ the dilaton decays primarily into $WW,ZZ$ pairs.  For $m_\chi<150~\rm{GeV}$, the $\mu^+\mu^-$ channel becomes comparable
to the $b \bar{b}$ channel. The dilaton mass and the muon Yukawa coupling
are correlated, so at large $m_\chi$ the $\mbox{BR}(\chi\to \mu^+ \mu^-)$
actually increases.  Even though our analysis has focused on the possibility of an enhanced $\chi \mu^+ \mu^-$ coupling, enhancements for the couplings to $e^+ e^-$ or $\tau^+ \tau^-$ can also lead to other clean discovery channels.   For $m_\chi<2 m_W$, these Yukawa couplings are more important than they are for a heavy dilaton which mainly decays into weak gauge bosons.

\begin{figure}[h]
\begin{center}
\includegraphics[width=1\linewidth]{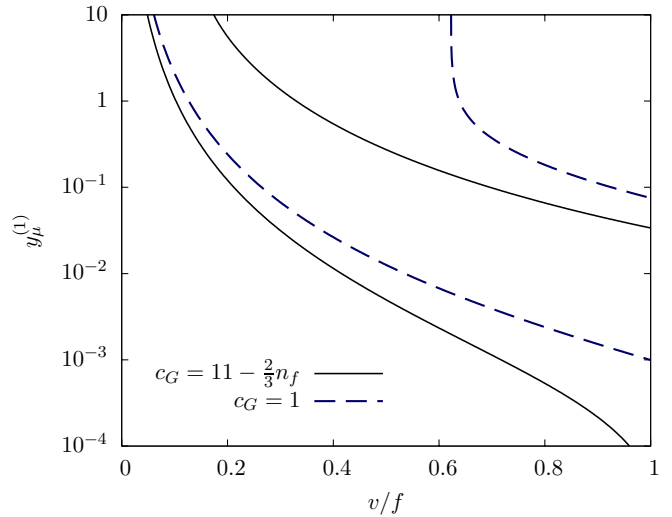}
\caption{Potential $5\sigma$ discovery regions in the $gg \to\chi \to \mu^+\mu^-$ channel for  a dilaton with $m_\chi = 150~\rm{GeV}$ with both ATLAS and CMS detectors collecting an integrated luminosity of $30~\rm{fb^{-1}}$ each.  For each value of $c_G$ shown, the discovery region lies between the two solid (dashed) lines. The lower curves represent bounds on the discovery region while the upper curves represent the Tevatron bounds from Fig. \ref{fig:tevee2}. \label{fig:discovery}}
\end{center}
\end{figure}

Finally, we present the $5 \sigma$ discovery range for $m_\chi=150~{\rm GeV}$
in the $(v/f,y_\mu^{(1)})$ plane, assuming an integrated luminosity ${\cal L}=30~{\rm fb}^{-1}.$   Ref.~\cite{Han:2002gp} (their Fig.~1) summarizes the magnitude of the modification to $\sigma(gg\rightarrow h\rightarrow \mu^+ \mu^-)$ due to new physics, $\kappa=\sigma_{New}/\sigma_{SM}|_{gg\rightarrow h\rightarrow \mu^+ \mu^-}$ required for both   $3
\sigma$ and $5 \sigma$ level discoveries with ${\cal L}=300 \rm{fb}^{-1}$ in both the ATLAS and CMS detectors at the LHC.    By rescaling their plot, we find that in our case a $5 \sigma$ discovery at ${\cal L}=30~\rm{fb}^{-1}$ requires $\kappa=17.4$  for $m_\chi=150~\rm{GeV}$.    The parameter space accessible at the LHC is presented in Fig.~\ref{fig:discovery}, taking into account bounds from direct searches at the Tevatron.

\section{Conclusions}
\label{sec:conclusions}
The breaking of electroweak symmetry could be 
triggered by the dynamics of a nearly conformal
sector. In such a case there will be a light
pseudo-Goldstone boson, the dilaton,  associated with 
the spontaneous breaking of conformal symmetry. 
We investigated properties of the dilaton, focusing
on possible deviations from the properties of the SM Higgs boson.

This scenario would be most interesting if the nearly
conformal sector is strongly interacting. 
However, few details of our analysis depend on the presence
of strong interactions. The non-derivative couplings
of the dilaton depend on the dimensions of operators that
explicitly break conformal invariance. These in turn depend on the
details of the underlying theory.

We considered contributions to the dilaton-fermion
couplings arising from  small explicit conformal symmetry breaking.
Since the Yukawa couplings in the SM are proportional to the
fermion masses, we studied the modifications of the Yukawa couplings
to the light fermions. Such couplings can be significantly enhanced
and this can lead to novel hadron collider phenomenology. Even modest
enhancements of the couplings to the light fermions are of great phenomenological relevance.
First, new production channels like $u \bar{u} \to \chi, d \bar{d} \to \chi$ may
become more important than the gluon fusion channel.
Second, new clean discovery signatures
like $\chi \to e^+e^-, \mu^+\mu^-$ can be viable unlike
in the case of the SM Higgs boson. These processes might provide
a handle on the dynamics at energy scales beyond the reach of the LHC.

\section{acknowledgments}
We thank A. Martin for valuable discussions.  This work has been supported in part by grant DE-FG-02-92ER40704
from the US Department of Energy. WG is also supported by the OJI
award.


\end{document}